\documentclass[chicago,numberedappendix,tighten]{emulateapj}
\usepackage{graphicx}
\usepackage{color}
\usepackage{ulem}

\shorttitle{Expanding Neutral Hydrogen Supershell}
\shortauthors{Chakraborti \& Ray}

\begin{document}

\title{An Expanding Neutral Hydrogen Supershell Evacuated by \\ Multiple Supernovae in M101}

\author{Sayan Chakraborti\altaffilmark{1} \& Alak Ray\altaffilmark{1}}
\affil{Department of Astronomy and Astrophysics, Tata Institute of Fundamental Research,\\
    Homi Bhabha Road, Mumbai 400 005, India}

\email{sayan@tifr.res.in, akr@tifr.res.in}


\altaffiltext{1}{Institute for Theory and Computation,
Harvard-Smithsonian Center for Astrophysics, 60 Garden St.,
Cambridge, MA 02138, USA}

\begin{abstract}
Several neutral hydrogen (HI) cavities have been detected in the Milky Way
and other nearby star forming galaxies. It has been suggested that at
least a fraction of them may be expanding supershells driven by the
combined mechanical feedback from multiple supernovae occurring in an
OB association. Yet most extragalactic HI holes have neither a demonstrated
expansion velocity, nor an identified OB association inside them. In this
work, we report on the discovery of an unbroken expanding HI supershell in the nearby
spiral galaxy M101, with an UV emitting OB association inside it. We measure
its size (500 pc) and expansion velocity (20 km/s) by identifying both its
approaching and receding components in the position-velocity space, using
21 cm emission spectroscopy. This
provides us with an ideal system to test the theory of supershells driven by the mechanical
feedback from multiple supernovae. The UV emission of the cluster inside the supershell is
compared with simulated spectral energy
distribution
of synthetic clusters
of the appropriate age ($\sim15$ Myr).
The observed UV flux is found to be consistent with an 
association of the appropriate mass ($\sim10^5 M_\odot$) and age required by the energy
budget of the supershell.
Properties of this supershell and another previously
reported in the same galaxy are used to infer its neutral hydrogen scale height
and mean neutral hydrogen density in the disk.
The presence of another UV emitting stellar association in over-dense swept up
gas is discussed in the context of propagating star formation.
\end{abstract}


\section{Introduction}
OB associations and other
young stellar clusters containing a few to thousands
of massive stars of spectral class O and B; harbor the majority of
massive stars in spiral galaxies. Stars of spectral class O are short-lived, and 
end their lives as supernovae within a few millions of years. These associations
last for few tens of millions of years because all their 
massive stars exhaust their nuclear fuel within this time.
Energetic radiation from OB associations ionize the interstellar medium (ISM) in 
galaxies and produce localized HII regions as well as diffuse ionized gas.
Supernovae of various types explode from the massive stars that form
in these associations which in turn pressurize and churn the
ISM \citep{2002AJ....123..255O}.
Expanding neutral hydrogen structures, called supershells were identified
in the Milky Way by \citet{1979ApJ...229..533H}.
Supershells form and evolve due to the continuous mechanical energy injection by stellar 
winds and SNe in massive parent OB associations \citep{1987ApJ...317..190M}. The outer
shocks of the  superbubbles  sweep up the ambient ISM 
into a thin cool shell.

In this work we report yet another striking case of an expanding HI supershell in
M101 which we argue is driven by multiple supernovae from a young stellar association.
We use the observed properties of the HI supershell to determine its dynamical age
and energy requirement.
Here we demonstrate that this HI cavity has measured expansion velocity
and at the same time is fully contained,
i.e.\ it has not yet broken out of the
disk of its host galaxy.
We also detect a UV emitting stellar association inside the
HI hole. We model its spectral energy distribution,
test and confirm the hypothesis of
\citep{1987ApJ...317..190M} that some supershells are driven by the combined action
of multiple supernovae occurring in a stellar association inside it.

These supershells may also play a role in the process of star formation and favor 
propagating star formation, as pointed out by
\citet{1987ApJ...317..190M,1994MNRAS.270...75P}. 
\citet{2008ApJ...682L..13H}, point out that the neutral and molecular
gas replenishment in the walls of the supershells may provide the trigger
for collapse and further star formation. In NGC 300, \citet{1997ApJS..108..261B}
reported several cavities, much larger than galactic supernova remnants (SNRs).
\citet{2004A&A...425..443P} suggest that these may represent the above mentioned supershells.

Cavities have been found in neutral hydrogen (HI) surface density maps of several
nearby galaxies (\citealp[for a classical review and inventory of HI holes,
see][]{1988ARA&A..26..145T},
\citealp[for a recent review and survey of HI holes, see][]{2010arXiv1008.1845B});
including M101 \citep{1991A&A...244L..29K}.
However, the presence of a cavity alone is insufficient evidence for the existence
of a supershell. It may simply be a low density region between higher density ones,
created by turbulent distribution of gas.
This pitfall is illustrated by the case of Holmberg II. In that galaxy
\citet{1992AJ....103.1841P} used underdensities in HI column density map to create a
catalog of HI cavities and
postulated that they have been evacuated by multiple SNe. However \citet{1999AJ....118..323R}
rule out a multiple supernova origin for most of these holes since they were unable to find
any young cluster or even trace of recent star formation activity which would have led 
to the SNe responsible for these HI-holes.
\citet{2000ApJ...529..201S} have since then re-examined the
\citet{1999AJ....118..323R} results and concluded that the energy available from
massive stars inside HI shells indicates that energy deposited into the ISM from
supernovae and stellar winds is sufficient to account for the HI morphology.
\citet{2006A&A...448..123S} have found that the observed properties of the most prominent
kpc-scale structure in IC 1613 and the level of the detected star formation activity are
inconsistent with the hypothesis that they are formed by energy injection from
multiple supernovae.

The lack of direct determination of the
source of energy driving these bubbles in many cases have led to numerous
suggestions regarding their
origin. \citet{1998ApJ...503L..35L} have suggested that a fraction of HI supershells
might be driven by Gamma Ray Bursts (GRBs). 
\citet{2000ApJ...540..797W} have discussed the possibility of forming cavities
from the nonlinear growth of thermal and gravitational instabilities in 
the ISM.
\citet{1981A&A....94..338T} and \citet{1984ApJS...55..585H} have suggested
that the impact of small companion galaxies or high velocity clouds with the disk of a
larger galaxy may form these cavities.
\citet{1982ApJ...253..666E} have proposed that radiation pressure from
field stars may provide a driving mechanism for supershells. \citet{1996ApJ...467..666O}
has provided a comprehensive comparison of the supernova driven supershell model
and the observed properties of the LMC shells.
Outstanding issues in the field are listed by
\citet{2007IAUS..237..106O}.

The first unambiguous case of an extragalactic expanding HI shell
associated with a hole in the HI surface density was reported by \citet{1991A&A...244L..29K}
in observations of M101 using the Westerbork Synthesis Radio telescope.
\citet{2009ApJ...691L..59W} have investigated the creation of a supergiant HI shell in
a dwarf irregular galaxy IC 2574 and its role in triggering star formation around it.
Star formation studies in most nearby galaxies indicate that sites of recent star formation
are correlated with sites of higher HI surface densities\citep{1998ARA&A..36..189K},
leading to several recipes for star formation.
The feedback processes described above should however clear out the
gas and lead to young stellar associations in regions of low gas density surrounded by 
higher density swept up shells. Therefore we began searching for cavities in the HI
surface density, occupied by young clusters.

\section{Data Description}
In this work we have used 21 cm line data to identify and measure kinematic
properties of our putative supershell. We have further used UV data to identify
the young stellar association responsible for driving it. We discuss below, the 
detailed nature of the data; especially the sensitivities, spatial and spectral
resolutions of the two relevant surveys.

\subsection{The HI Nearby Galaxy Survey}

The HI Nearby Galaxy Survey (THINGS) undertaken at the NRAO
Very Large Array (VLA) to study the HI emission from nearby galaxies
\citep{2008AJ....136.2563W} provides high angular resolution
($\sim7''$) maps of HI emission and high spectral
resolution ($\sim5$km/s) data. The survey has acquired radio interferometric
observations for the target galaxies and presents the data in the form of
image cubes generated from the interferometric data after background
continuum subtraction.
The three axes in these cubes represent the Right Ascension (RA),
Declination (Dec) and recession velocity. For each triad of these coordinates,
the cube gives the HI line flux density, in excess of the best estimate
for the continuum. Information about the position
and velocity of neutral hydrogen gas can be extracted from this data cube. 

Demanding that the diameter of the supershell
be more than the projected beam size, we can only detect supershells with radii more than
$24\mathtt{\; pc \;}(\theta_{10}D_1)$,
where $\theta_{10}$ is the beam size in units of 10'' and $D_1$ is the distance to the host
galaxy in Mpc. The two velocity components can be resolved only if the velocity is greater than
the spectral resolution.
This is of course limited by the sensitivity to HI in the beam, which is about
$\sim10^{20}$ atoms/cm$^2$ for the data being used.

\subsection{GALEX Nearby Galaxies Survey}

Data on M101 from Far and Near Ultraviolet
observations are available from GALEX Nearby Galaxies Survey, conducted with the Galaxy
Evolution Explorer (GALEX) \citep{2005ApJ...619L...1M}. The ultraviolet emission provides us
with a simple tool for identifying young stellar associations, which are expected to
host a sizable fraction of 
the core collapse SNe, which supply the energy budget for the expanding supershells.
We use information about both the HI distribution and the young stellar populations,
to search for neutral hydrogen supershells powered by multiple supernovae.

The GALEX data is provided as FITS files where each pixel gives the counts per second (cps),
in the relevant band,  for each RA-Dec pair.
Comparing the HI surface density maps with those for near ultraviolet (NUV) emission for M101,
reveals multiple regions with high UV emission but low HI surface density, surrounded by
regions of higher HI density. These include the HI superbubble reported by Kamphuis
et al.(1991). Another association hosting SN 1970G, is detected in the ultraviolet and
found to be located on the high column density HI ring surrounding an underdensity.
We have also found a separate cavity in the HI surface density,
hosting a young stellar association of NUV magnitude $m_{AB}=17.37\pm0.08$,
surrounded almost completely by higher HI surface density
regions (Fig. 1). This region, which shall be discussed further in this work, is
located approximately at 14h 02m 25s +54d 27' 45''.

\section{Evidence for expansion}

Most extragalactic HI cavities have no demonstrated expansion velocities
(see exceptions in Introduction) Demonstration of expansion
velocities is essential for the identification of HI cavities associated
with supershells.
Even where
expansion is claimed, the measured velocity is sometimes
close to the typical rms
velocity. In absence of clear detections for both forward and backward velocities,
claims for expansion are ambiguous. For example \citet{2009ApJ...691L..59W} infer
a velocity from a break in the P-V diagram, which they claim to be an indirect
measure of expansion, yet consistent with presently stalled expansion.
Because of sensitivity limitations, the velocity profile of the HI gas in an
underdense region becomes difficult to measure. \citet{1992AJ....103.1841P} point
out that because of the low brightness contrast between the shells and the face of
their HI-holes, they find it very difficult to look for expansion
in position velocity (P-V) diagrams and to demonstrate both incoming and receding
components. The high sensitivity of the THINGS survey, used in our study, overcomes
this limitation.

To study the dynamics of the HI gas, we extract the relevant region (entire area covered in  Fig. 1)
out of the THINGS data cube and transform it, using the Astronomical Image Processing System
(AIPS http://www.aips.nrao.edu/),
to obtain velocity information on the sky plane.
We slice the resulting data sub-cube along nine equally spaced Dec values over the
entire region of interest (whole of Fig. 1). The HI line intensity is then plotted in the
P-V diagrams for these slices (Fig. 2) The P-V diagrams have a spatial resolution of
10'' and a velocity resolution of 5 km/s.
The HI distribution in the starting and ending slices (Fig. 2) are continuous over the RA
and velocity, indicating that most of the gas is at the local rest frame velocity dictated
by the galaxy rotation curve. The Dec slices cutting across the HI underdensity however indicate
a clear transition to a two-component HI emission.
The true nature of the HI-hole can however only be studied from its detailed velocity map. 
The data slice along Dec slice D clearly shows that the HI under-dense region is characterized by
reduced emission from the rest frame velocity of the HI disk.
This provides an
unambiguous detection of an expanding HI supershell surrounding a UV emitting stellar association.
The difference in velocity of the two components in the position velocity diagram (Fig. 2
panel D) is around 40 km/s. Interpreting this as twice the expansion velocity of the
supershell, we get $V_S\sim20$ km/s. The diameter of the cavity, which lacks HI emission
at the rest frame velocity of the disk, is around $\sim30''$, corresponding to
$R_S\sim500$ pc at the distance of 6.85 Mpc to M101 reported by \citet{2006ApJS..165..108S}.
We also note the bilateral symmetry in the emission from the two detected components and suggest
that the supershell must have started out at or near the mid-plane of the HI disk of the host
galaxy.

We also overlay the HI column density contours
on a map of rms deviations of velocities from the mean, obtained by taking the second moment
of the velocity distribution of the emission at each pixel (Fig. 3). The square of the rms
velocities trace the line of sight kinetic energy per unit mass of the HI gas.
Hence the high
rms velocity of up to 18.1 km/s in the HI-hole region is evidence for recent energy injection
into the ISM. For comparison, the neighborhood of the HI-hole, reveals an average rms velocity 
of 10.5 km/s with a standard deviation of 1.4 km/s.
We construct a continuum subtracted HI spectra (Fig. 4), integrated over a box
(2.6$\times$beamsize on each side) inside the HI-hole. 
It clearly shows that there are two components of HI emission with easily
differentiated velocities. We interpret the two components, as demonstrated in Fig. 2 and
Fig. 4, as the approaching and receding
hemispheres of an expanding HI supershell around an evacuated bubble.
Note that the sensitivity of the THINGS survey allows us to detect the gas even in the
lowest density region with high significance.

\section{Implications for Host Galaxy}

HI observations of face-on galaxies map their neutral hydrogen surface densities.
However HI scale heights have been directly measured mostly for edge on systems.
Scale height measurements for face-on galaxies are rare.
Here we propose to put constrains on the HI scale height of M101 based on the
properties of the described supershell and the one reported by \citet{1991A&A...244L..29K}.
\citet{1980ApJ...238L..27B} have discussed the sizes of galactic HI shells as a function of
the neutral hydrogen scale height.

\subsection{Supershells in M101}

\citet{1991A&A...244L..29K} have argued on the basis of the size of the hole found by them
and the velocity width of its HI profile that it has already reached a state where
parts of it are accelerating away from the disk and into the halo. At the quoted
diameter of 1.5 kpc, if the supershell has burst out of the galactic HI disk, the HI scale
height is likely to be less than $R_S\sim750$ pc. On the other hand both the approaching
and receding components of the HI profile for the hole reported in our work (Fig. 2 panel D)
are continuous in position and velocity (unlike Fig. 3 in \citet{1991A&A...244L..29K})
and appears not to have burst out of the HI disk. Therefore the scale height
is likely to be more than $R_S\sim500$ pc.

\subsection{Scale Height of M101}

\citet{1987ApJ...317..190M} explain that the pressure
driven phase of a supershell terminates when $R_S$ becomes comparable to the HI scale height.
Thereafter, if the velocity of the shell at one scale height
is greater than the typical RMS velocity of the HI,
portions of the shell traveling perpendicular to the disc accelerate into the
halo, fragment due to Rayleigh-Taylor instabilities and discharge the internal pressure
into the galactic corona. The portion moving
along the disc, enters the pressure-less snowplow phase and continues to evolve
slowly ($R_S \propto t^{1/4}$). Hence, the radii of the largest observed HI-holes in
face-on disk galaxies are expected to be slightly larger than the HI scale height.

Observations of these two (\citet{1991A&A...244L..29K} and this work) supershells
constrain the HI scale height of M101 to be between 500 pc and 750 pc.
The average line integrated flux, from the region covered in Fig. 1, is 0.18
Jy/beam $\times$ km/s. We use the conversion from \citet{2000tra..book.....R} to get an
average HI column density of $N_{HI}\sim1.8\times10^{21}$ atoms/cm$^2$.
Even the least dense area in Fig. 1 has $N_{HI}\sim0.35\times10^{21}$
which is well above the detection threshold.
Using a scale height of 600 pc consistent with the above constraints
we infer the initial gas density by dividing the column density by twice the scale
height to get $n_0\sim0.5$ in units of atoms cm$^{-3}$. We have used
this value while
estimating the energy budget of the supershell later in this work.

\section{Inversion of McCray \& Kafatos Model}

Massive stars in a young stellar association end their lives as core collapse
supernovae. In an OB association this
may result in the formation of an expanding neutral hydrogen supershell driven by the
mechanical feedback from the combined action of multiple supernovae.
\citet{1987ApJ...317..190M} demonstrate that the rate of SNe from a cluster of typical
IMF, varies only slightly over the first 50 Myr and can be thought of as a continuous
energy injection over such a timescale, powering the expansion of the supershell.
This provides a particularly simple framework, which we adopt, for understanding the evolution
of a supershell, pressure-driven by feedback from multiple supernova. The radius ($R_S$) and
velocity ($V_S$) of an expanding supershell, given by them, are
\begin{equation}
 R_S=97\mathtt{\; pc \;}(N_*E_{51}/n_0)^{1/5} \; t_7^{3/5}
\end{equation}
and
\begin{equation}
 V_S=5.7\mathtt{\;km\;s}^{-1}\;(N_*E_{51}/n_0)^{1/5} \; t_7^{-2/5}.
\end{equation}
Where, $N_*$ is the number of massive ($>7M_\odot$) stars in the association responsible for
creating the supershell. $E_{51}$ is the energy produced by each supernova explosion.
$n_0$ is the assumed uniform atomic density prior to the formation of the supershell and
$t_7$ is the age of the supershell in units of $10^7$ years.
For an alternate formulation
of the problem, which directly relates the evolution to the supershell to the
mechanical luminosity of the driving cluster, see Appendix.

The \citep{1987ApJ...317..190M} model is invertible in principle, because the $R_S$
and the $V_S$ are directly observable from the HI data. In this work, we therefore
re-frame the equations as
\begin{equation}
 t_7 = (R_S / 97\mathtt{\; pc}) (V_S /5.7\mathtt{\;km\;s}^{-1})^{-1}
\end{equation}
and
\begin{equation}
 (N_*E_{51}/n_0) = (R_S / 97\mathtt{\; pc})^2 (V_S /5.7\mathtt{\;km\;s}^{-1})^3
\end{equation}
to express the variables, which characterize the supershell, purely in terms of the observable
quantities. These equations will allow us to estimate the age and
mass
of an association which will suffice to reproduce the observed size and expansion
velocity of any given supershell. If there is no young stellar association interior
to an observed expanding neutral hydrogen structure or an association smaller than
what is demanded by these set of equations, then the \citep{1987ApJ...317..190M}
model can be ruled out. If on the other hand, one finds such a sufficiently
luminous association, the model can then explain the origin of such a supershell.

\section{Age and energy budget of the M101 supershell}

\citet{2009ApJ...691L..59W} have used the model of \citet{1974ApJ...188..501C},
to compute the energy required to evacuate the HI mass inside the supershell
that they study in IC 2574. Chevalier's model deposits the total energy at
the start of the simulation and he notes that this soon relaxes
to an approximation of the Sedov solution. This was intended for simulating a single
SNR. However this is not a suitable description of the energy
input from multiple SNe as stars of different initial masses, explode at different points
in time. In addition note that the power law index of time evolution (3/5 in Eq.1),
due to continuous energy injection, is different from the Sedov-like solution (used by
\citet{2009ApJ...691L..59W}) and is similar to the wind solution of
\citet{1977ApJ...218..377W}.

The appropriate evolution of the supershell can be obtained by the
 self consistent solution
of Eqs (1) and (2). The observed size and expansion velocity of the supershell will be 
satisfied in this manner for
\begin{equation}
 t_7 \sim 1.5
\end{equation}
and
\begin{equation}
 (N_*E_{51}/n_0) \sim 1.1 \times 10^3.
\end{equation}
Assuming $E_{51}\sim1$ and $n_0\sim0.5$ (as already demonstrated), we need the
driving cluster to be massive enough
to harbor $N_* \sim 550$ supernova yielding massive stars. Assuming the
total mass of stars required to be formed to produce each core
collapse supernova as 196 $M_\odot$ following \citet{1997ApJ...476..144M}, we need
a cluster of initial mass $\sim 10^5M_\odot$ which has been driving this
supershell into the ISM for the last $\sim15$ Myr. Only $\sim15\%$ of the
energy from the already exploded supernovae (at $t_7=1.5$), needs to be in the kinetic
energy of the supershell, assuming $n_0=0.5$.
This is close to the estimate from \citet{1987ApJ...317..190M}. According
to Starburst99 \citep{1999ApJS..123....3L} the number of ionizing photons with wavelength
below 912\AA, drops to $4\times10^{-4}$ times the initial rate, by the time a typical
association reaches 15 Myr. Hence, given the estimated age of the supershell
it is no surprise that it is made up of neutral hydrogen.

\section{UV diagnostic for mass of the cluster}

Most supershells in the Milky Way suffer from large extinctions
\citep{1988ARA&A..26..145T}. Even for many extragalactic holes, young
stellar populations have not been detected inside \citep{1999AJ....118..323R}.
Using Starburst99 with the parameters as described in \citet{1999ApJS..123....3L}
we simulate the spectral energy distribution (SED)
of an instantaneous starburst
(Salpeter IMF with $\alpha=2.35$ between $1-100 M_\odot$) of solar metallicity
($Z=0.02$) aged 15, 50 and 200
Myr, over a range of wavelengths from $500\AA$ to $3500\AA$ (see Fig. 5). The
effective collecting area of the GALEX NUV and FUV are plotted on the same graph
for comparison. At the effective wavelength of the GALEX NUV band the luminosity
of the 15 Myr (chosen to match the dynamical age of the supershell) starburst is
given as $1.17\times10^{32}$ ergs sec$^{-1}$ $\AA^{-1}$ $M_\odot^{-1}$. 
It is to be noted, that as an association ages, over the time range represented in
the figure, the flux in both GALEX bands falls by more than one order of magnitude.
This motivates the use of GALEX images in locating \textit{young} stellar association
in nearby galaxies, which may be the source of multiple supernovae driving
supershells as suggested by \citet{1987ApJ...317..190M}.

In this work, we measure the NUV emission of the cluster interior to the
HI cavity using archival GALEX observation. We find the NUV magnitude,
uncorrected for extinction, to be $m_{AB}=17.37\pm0.08$. Using only the galactic
extinction of $A_V=0.028$ from \citet{1998ApJ...500..525S}, and scaling it to the
NUV band using the relationship from \citet{1989ApJ...345..245C}, we get
$A_{NUV}=0.083$, for a typical Milky Way ratio of total to selective absorption
of $R_V=3.1$. At the adopted distance to M101, this gives an NUV luminosity of
$1.51\times10^{37}$ ergs sec$^{-1}$ \AA$^{-1}$.
Comparing the simulated SED from Starburst99 (as above) and the observed GALEX flux,
we infer the mass of the putative young stellar association responsible for evacuating
the observed supershell. 
Assuming that the age of association is same as the dynamic
age of the supershell as determined in this work,
the association had an original mass of $\sim 1.3\times10^5M_\odot$.
This is consistent with the minimum mass requirement of $\sim 10^5M_\odot$ that
was derived from the energy requirement in the \citet{1987ApJ...317..190M} model.
The inferred mass of this cluster is consistent with masses of the most massive
of young stellar clusters (or super star clusters) seen in the local universe
\citep{2003dhst.symp..153W}.
Thus we find that the \citet{1987ApJ...317..190M} model provides a simple and
sufficient explanation of the observed supershell in M101.

\section{Instabilities of the Expanding Supershell}
Apart from the NUV emission from stars inside the bubble, we also note the even stronger 
NUV emission from the high density swept up ring to the south of the cavity (Fig. 6)
at around 14h 02m 25s +54d 27' 15''.
We suggest that this is evidence for induced star formation at the edges of
expanding supershells driven by multiple supernovae from an OB association.
\citet{1989ApJ...340..786E,1994ApJ...427..384E} has investigated the gravitational
collapse of decelerating shocked
layers in OB associations and proposed a condition for propagating star
formation due to large expanding shells. \citet{1994ApJ...428..186V} has presented
hydrodynamical stability analysis for cold gas bounded by shocks.

\citet{1997A&A...328..121E} performed a linear analysis of the
\textit{Elmegreen-Vishniac} instability and
derived an instability time ($t_b$) which marks the onset of
gravitational instability of a self gravitating expanding supershell. Re-casting their
expression in the notation of \citet{1987ApJ...317..190M},
\begin{equation}
 t_b = 47\mathtt{\; Myr \;} (N_* E_{51})^{-1/8} c_1^{5/8} n_0^{-1/2} \mu_{1.3}^{-1/2}
\end{equation}
where $c_1$ is the sound speed in the shell in units of 1 km/s and $\mu_{1.3}$ is the mean
molecular weight in units of 1.3 amu. Using $N_* \sim 5.5 \times 10^2$, $n_0\sim0.5$ and unity
for the rest of the dimensionless quantities, we get $t_b \sim 30$ Myr.
Comparing this with the estimated age of the supershell (Eq 5),
we find that the supershell is not yet prone to
gravitational instabilities leading to star formation.
Since there is star formation at the boundary of the region discussed before,
this then may be due to the expanding supershell encountering an existing
dense cloud triggering star formation (\citet{1998ASPC..148..150E} and references
therein). Such a scenario can also provide a natural explanation for the deviation
from spherical symmetry in the HI-hole.

\section{Discussions}
We have reported the discovery of an expanding HI supershell in M101, which hosts a
young stellar association in its cavity.
The supershell may have been created by the combined effect of multiple supernovae.
This allowed us to use the McCray \& Kafatos model to estimate the
age and mechanical luminosity required from the cluster
that harbored the massive stars responsible for these explosions.
By comparing the observed GALEX NUV emission of the cluster found inside the HI-hole,
with the synthetic SED of the required age obtained using Starburst99,
we demonstrate that the cluster
is heavy enough to host enough massive stars which explode as supernovae and
provide energy to drive the supershell.
We have discussed the implications of the measured
properties of the supershell in the host galaxy to estimate the galaxy's HI scale height
and the mean neutral hydrogen density in the disk.
We also note the presence of young stellar associations
near the edge of the observed supershell and discussed the role of instabilities
in propagating star formation.

Further studies, e.g. high resolution imaging or
integral field optical spectroscopy of the cluster in the HI-hole, may determine
the metallicity enrichment of the gas and through the use of spectral fitting of
stellar populations, allow the determination of true age and initial mass of the cluster.
Spatially resolved integral field spectroscopy also has the potential to
reveal the role of multiple stellar populations, if any, within the cavity. 
We have shown that the observed UV fluxes of the cluster are consistent
with the standard model of supershell formation. An independent determination of the
age of the cluster will prove the applicability of the model in this case.
Sensitive X-ray observations may reveal emission from possible hot coronal gas inside the
supershell \citep{2001MNRAS.324..191S,2005ApJ...635.1116S}.
However, \citet{1995ApJ...450..157C} suggest that an intermittent process
such as hidden supernova remnants may be responsible for the X-ray-bright super bubbles 
and hence, not all superbubbles are X-ray bright.

In case of triggered star formation, \citet{2000ApJ...529..201S}
have emphasized the importance of establishing that stars are formed in chains
of progressing age.
We predict that spectroscopy will reveal the populations noted near
the edge of the supershell, to be younger than the cluster seen
inside the supershell.
A detailed study of this supershell with the upcoming ALMA may render CO line
maps which will trace the cold molecular gas at the sites of triggered star formation.
\citet{2010arXiv1008.1845B} have now produced a catalog of more than 1000 HI holes
in 20 nearby galaxies from the THINGS sample. A lot of these galaxies
have deep GALEX observations, which we have demonstrated as a simple tool for
identifying possible young stellar associations supplying the energy requirement.
This could form the basis of a much larger study to determine the fraction of
HI holes that can be explained by the \citet{1987ApJ...317..190M} model.

\acknowledgments

We thank Richard McCray for discussions at Darjeeling which motivated this work.
We thank an anonymous referee for detailed comments which helped in improve this work.
SC thanks Nimisha Kantharia and Nirupam Roy for discussions on AIPS methods for handling
HI data.
This work has made use of data from The GALEX Nearby Galaxies Survey. The Galaxy Evolution
Explorer (GALEX) is a NASA Small Explorer, launched in April 2003.
This work made use of THINGS, The HI Nearby Galaxy Survey, conducted with the NRAO Very Large
Array (VLA). We thank the Institute for Theory and Computation, Harvard University for
its hospitality. At Tata Institute of Fundamental Research, this work is
supported by the 11th Five Year Plan 11P-409.


\clearpage

\begin{figure}
\includegraphics[angle=0,width=0.9\columnwidth]{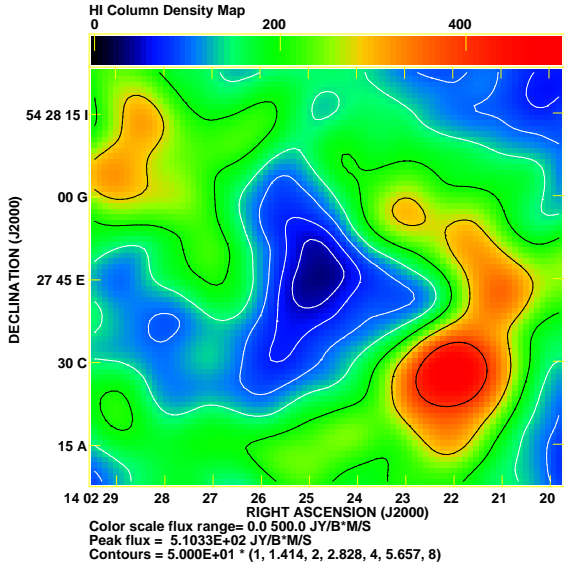}
\caption{HI surface density map of M101 at the location of the cavity surrounded by
higher surface density of gas. Note the beam size is 10'', approximately the same as
the central under-dense contour. Alphabets beside the declination
markings, denote the positions of the declination slices in Fig. 2.}
\end{figure}


\begin{figure}
\includegraphics[angle=0,width=0.9\columnwidth]{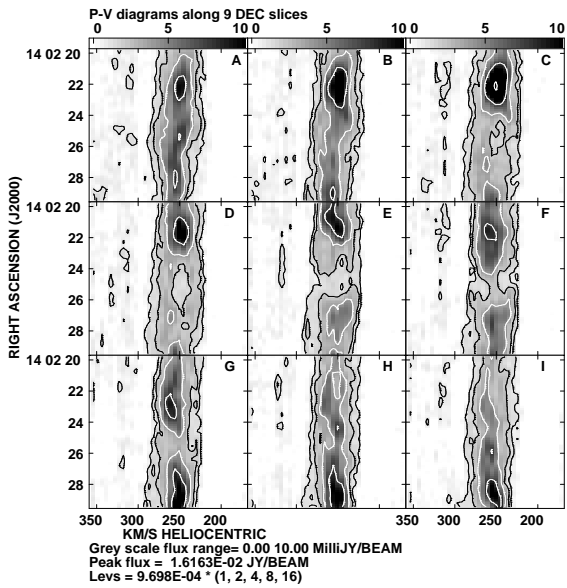}
\caption{HI gas velocity distribution vs RA in nine equally spaced Dec slices across the
region shown in Fig. 1. Note the two resolved velocity components surrounding the HI-hole,
e.g. in panels C and D.}
\end{figure}


\begin{figure}
\includegraphics[angle=0,width=0.9\columnwidth]{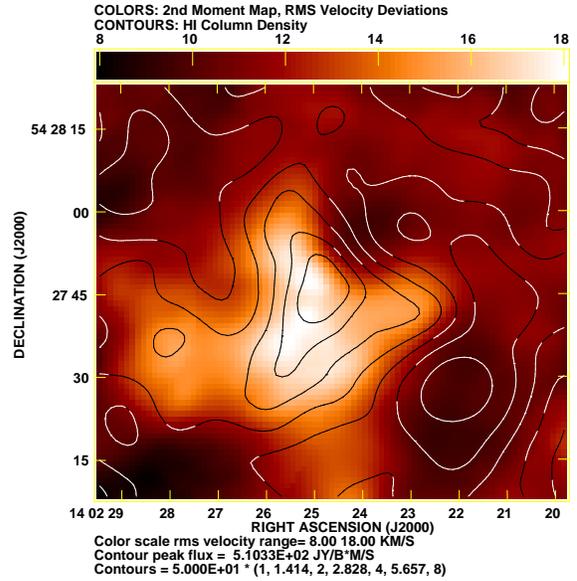}
\caption{HI surface density contours, plotted over the 2nd moment of the HI velocity
distribution map (RMS deviations in velocity from the mean) in color. Note that the
highest kinetic energy per unit mass are observed within the HI-hole.}
\end{figure}


\begin{figure}
\includegraphics[angle=0,width=0.9\columnwidth]{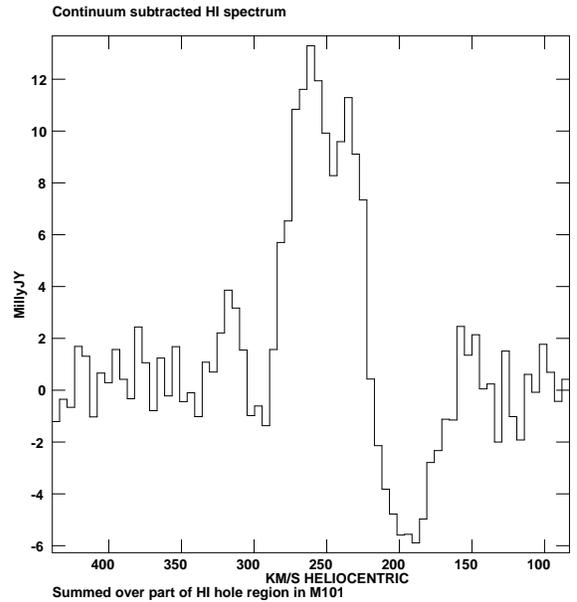}
\caption{Continuum subtracted HI spectrum integrated over a box,
2.6$\times$beamsize on each side, inside the HI-hole. Note the
double peaked emission profile.}
\end{figure}


\begin{figure}
\includegraphics[angle=0,width=1.0\columnwidth]{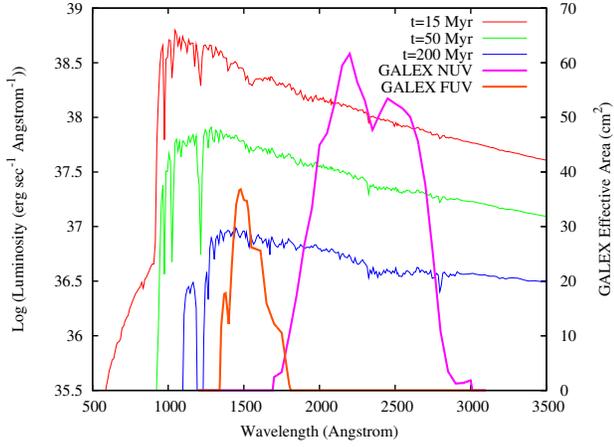}
\caption{Spectral Energy Distribution (SED) of a $10^6 M_\odot$ association
(with Salpeter IMF and solar metallicity) at the age of
15, 50 and 200 Myr computed using Starburst99 \citep{1999ApJS..123....3L},
plotted for comparison with the effective collecting area of
GALEX satellite telescope in the NUV and FUV bands. Note that the flux in both
bands falls by more than an order of magnitude in the depicted range of ages.}
\end{figure}


\begin{figure}
\includegraphics[angle=0,width=0.9\columnwidth]{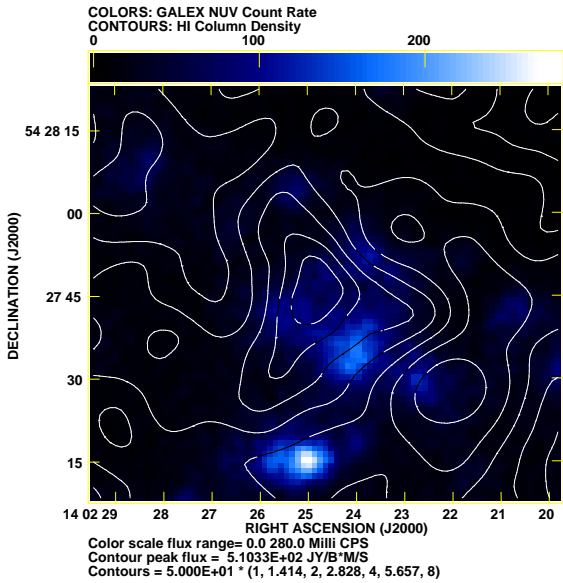}
\caption{HI surface density contours, plotted over the GALEX NUV photon count rate
(in color). GALEX typically produces images with 5" FWHM. Note the cluster inside the
HI-hole and also the brighter one on the swept up shell in the south.}
\end{figure}

\clearpage

\bibliographystyle{apj}
\bibliography{M101}

\appendix

\section{Supershell Dynamics: Alternate Formulation}

In this work we have used the early formulation of the supershell dynamics as
given by \citet{1987ApJ...317..190M}. This formulation depends directly on terms like
$N_*E_{51}$ which relate the dynamics to the total number of supernova yielding
stars in the associations. However in a later work,
\citet{1988ApJ...324..776M,1995RvMP...67..661B}
have directly linked the evolution of the supershell with the combined mechanical
luminosity of the stellar winds and supernovae from the association. 
\citet{1988ApJ...324..776M} give the radius and velocity evolution of the
supershell as
\begin{equation}
 R_S=267\mathtt{\; pc \;}(L_{38}/n_0)^{1/5} \; t_7^{3/5}
\end{equation}
and
\begin{equation}
 V_S=15.7\mathtt{\;km\;s}^{-1}\;(L_{38}/n_0)^{1/5} \; t_7^{-2/5},
\end{equation}
where the new term $L_{38}$ is the mechanical luminosity of the cluster in units
of $10^{38}\mathtt{\;erg\;s}^{-1}$. The dynamics is identical to the
\citet{1987ApJ...317..190M} model and hence can be inverted to give,
\begin{equation}
 t_7 = (R_S / 267\mathtt{\; pc}) (V_S /15.7\mathtt{\;km\;s}^{-1})^{-1}
\end{equation}
and
\begin{equation}
 (L_{38}/n_0) = (R_S / 267\mathtt{\; pc})^2 (V_S /15.7\mathtt{\;km\;s}^{-1})^3,
\end{equation}
which provide the required input parameters as a function of the observed
properties of the supershell. The expression for the age is identical to that
derived earlier in this work. Substituting the values for $R_S$, $V_S$ and $n_0$,
the required mechanical luminosity for the cluster
is $3.6 \times 10^{38}\mathtt{\;erg\;s}^{-1}$.

This mechanical luminosity may now be compared directly with those obtained from
population synthesis models. For the fiducial Starburst99 \citep{1999ApJS..123....3L}
simulation described earlier in this work, the mean mechanical luminosity in the first
15 Myr from stellar winds and supernovae is given as
$2.57\times10^{34}\mathtt{\;erg\;s}^{-1}M_\odot^{-1}$. Comparing this with the
energy budget, requires a massive $\sim1.4\times10^4 M_\odot$ young stellar cluster.
But, this is lower than
the mass required in the earlier analysis and hence the observed UV flux is consistent
with a cluster massive enough to drive the supershell.

\end{document}